\documentstyle[epsf]{neu}
\begin{document}

\title{%
EQUATION OF STATE AND OPACITIES  
FOR HYDROGEN ATMOSPHERES 
OF STRONGLY MAGNETIZED 
COOLING NEUTRON STARS
}

\author{Alexander Y.\ POTEKHIN, Yuri A.\ SHIBANOV \\
{\it    Department of Theoretical Astrophysics, 
Ioffe Physical-Technical Institute,
St.-Petersburg 194021, RUSSIA,
palex@astro.ioffe.rssi.ru}\\
Joseph VENTURA\\
{\it Physics Department, University of Crete, 
	  and Institute for Electronic Structure and Laser, FORTH, 
	   71003 Heraklion, Crete, GREECE}}

\maketitle

\section{Introduction}

Thermal radiation from surfaces of several radio pulsars   
has been  detected 
\linebreak
recently in soft X-rays by {\it ROSAT\/} and 
{\it ASCA\/} (see Becker and Tr\"umper,~1998). 
A number of the point-like X-ray sources has also 
been discovered and identified as radio-silent 
isolated cooling neutron stars (NSs) 
(see Caraveo et al.,~1996; Walter et al.,~1996; 
Vasisht et al.,~1997). In several cases, the observations seem to be 
better explained if the NSs possess an envelope 
of matter of low atomic weight,  
presumably hydrogen (Page,~1997), 
and if NS atmosphere models are applied to fit the data 
(Pavlov and Zavlin,~1998). 
There are indications that this may be the case for 
NSs with  different 
magnetic fields $B$: from weak, $B\ll10^{10}$~G 
(Rajagopal and Romani,~1996; 
Zavlin and Pavlov,~1998; 
Zavlin et al.,~1998), 
to strong, $B\sim10^{11-13}$~G (Page et al.,~1996; 
Zavlin et al.,~1998), 
and superstrong, $B>10^{14}$~G 
(Heyl and Hernquist,~1997). 
To justfy this, one needs 
in further observations and 
advanced atmosphere models of NSs 
with high $B$. 
Thermal motion of atoms 
accross the magnetic field 
induces an electric field  
in the frame of the atom. 
This affects atomic structure, 
atmospheric thermodynamics and  
opacities (Ventura et al.,~1992; 
Pavlov and M\'esz\'aros,~1993), being a 
critical point in the 
construction of  
models and  
data interpretation.   
Here we study this problem 
 for hydrogen plasma 
at temperatures $T\sim 10^{5.5}-10^{6}$~K, 
densities $\rho\sim10^{-3}-10^1{\rm~g~cm}^{-3}$, 
and magnetic fields $B\sim 10^{12}-10^{13}$~G, 
typical for atmospheres  of middle-aged cooling NSs. 
We construct an analytic model of the plasma free energy  
and derive a generalized Saha equation which 
is used to obtain the opacities. 

\section{Method and results}  
We use a generalization of the method applied previously 
to the non-magnetic case (Potekhin, 1996a). 
The Helmholtz free energy of a volume $V$ of the plasma is 
$ 
F(V,T,\{N_\alpha\}) = F_{\rm id}^{(p)}+F_{\rm id}^{(e)}+
F_{\rm id}^{(H)}+F_{\rm ex}, 
$
where the subscripts ``id'' and ``ex'' refer to the ideal 
and excess part of $F$, 
($e$, $p$, $H$)  
relate to the electrons, protons, and atoms, 
respectively, and $\{N_\alpha\}$ 
is the set of numbers of the particles. 

For protons, 
$
\beta F_{\rm id}^{(p)}/N_p =
      \ln(2\pi a_m^2\lambda_pn_p)
    + \ln\left[1-{\rm e}^{-\beta\hbar\omega_{cp}}\right]-1,
$
where $\beta\equiv1/{k_B} T$, $n_\alpha\equiv N_\alpha/V$, 
$\lambda_\alpha
\equiv(2\pi\beta\hbar^2/m_\alpha)^{1/2}$, 
$a_m\equiv(\hbar c/eB)^{1/2}$, 
$\omega_{cp}\equiv eB/m_p c$.  
Being the same for free protons and atoms,
proton spin and zero-point energy do 
not affect thermodynamics, therefore the corresponding terms 
are suppressed.

For electrons, we take into account partial degeneracy and 
relativity: 
\linebreak
$
F_{\rm id}^{(e)}/V=\mu n_e - P_{\rm id}^{(e)}
$,  
where the pressure 
$   P_{\rm id}^{(e)}
$
is easily expressed through fitting formulae  
by Blinnikov et al. (1996), and $n_e$ is related to 
the chemical potential $\mu$ 
through a fitting formula by Potekhin (1996b). 
For $F_{\rm ex}^{(ep)}$, we use 
an analytic expression (Chabrier and Potekhin, 1998) 
based on calculations by Chabrier (1990) at $B=0$. 
The latter approximation is justified for the classical 
plasma (according to the Bohr-van Leeuwen theorem)
but fails in the region where $B$ 
suppresses strong electron degeneracy 
(that occurs only in subphotospheric layers). 

For atoms, 
$\beta F_{\rm id}^{(H)}=\sum_{s\nu}\int{\rm d}^2 K_\perp N_{s\nu}
(K_\perp)\left\{\ln\left[n_H\lambda_H^3w_{s\nu}(K_\perp)/Z_w\right]
-1\right\}$, 
$s$ 
\linebreak 
and   $\nu$ 
relate to electronic excitations,  
$N_{s\nu} = (\lambda_H/2\pi\hbar)^2
N_H w_{s\nu}{\rm e}^{\beta \chi_{s\nu}}/Z_w$ 
are the atomic occupancies 
per unit phase space of the transverse component $K_\perp$ 
of the 
\linebreak 
pseudomomentum ${\bf K}$ which characterizes the atomic motion
in the magnetic field, 
\linebreak
 $w_{s\nu}(K_\perp)$ and $\chi_{s\nu}(K_\perp)$ 
are the occupation probabilities  
and binding energies 
of the moving atom, and
$Z_w=(\lambda_H/2\pi\hbar)^2
\sum_{s\nu}\int {\rm d^2}{K_\perp}
w_{s\nu}({K_\perp})\exp[\beta\chi_{s\nu}(K_\perp)]$ 
is the internal partition 
function. Large values of $K_\perp$ 
correspond to the decentered states with large dipole moment 
created by the induced electric field. 
These states have 
\linebreak
been  neglected in previous studies   
(e.g., Lai and Salpeter, 1997) but our results 
\linebreak
show that they are important at $T\gsim 10^{5.5}$~K. 
Generalization of the 
unbinding free energy (Potekhin, 1996a) 
to the magnetic case gives  
$\beta F_{\rm ex}^{(H)}=(N_H+ N_e)
\linebreak
\sum_{s\nu}\int~{\rm d}^2{K_\perp}
N_{s\nu} v_{s\nu}/V$, where 
$v_{s\nu}({K_\perp})=(4\pi/3)(r_{s\nu}^{(ep)}({K_\perp}))^3$ 
is an effective interaction volume and 
$r_{ep}$ the root-mean-square electron-proton distance in the atom. 

Minimization of $F$ leads to the generalized Saha equation
\begin{equation} 
    n_H =                                             
    {\lambda_p\lambda_e{(2\pi a_m^2)^2 \over \lambda_H^3}}
    \, n_en_p 
    \, \left[1-\exp(-\beta\hbar\omega_{cp})\right] \, 
    Z_w \exp(\Lambda_{\rm deg}),
\end{equation}
where 
$\ln w_{s\nu}\approx
\beta (\partial F_{\rm ex}^{(ep)}/\partial N_p+
\partial F_{\rm ex}^{(ep)}/\partial N_e-
(n_e+n_H) v_{s\nu}({K_\perp})$, 
and $\Lambda_{\rm deg}=\beta [\mu +
\partial\mu/\partial \ln n_e-
\partial P_{\rm id}^{(e)}/\partial n_e ]
-\ln(2\pi n_e\lambda_e a_m^2)$ is
the correction for electron degeneracy. 
In numerical solutions of Eq.~(1) we use fitting 
formulae for $\chi_{s\nu}$  and $r_{s\nu}^{(ep)}$
(Potekhin, 1998). 
The equation of state (EOS) is calculated from  
$P=-(\partial F/\partial V)_{T,\{N_\alpha\}}$.
\pagebreak

\begin{figure}[t]
\vspace{-1.5cm}
\begin{center}
\leavevmode
\vspace{.7cm}
\epsfxsize=4.cm
\epsfbox{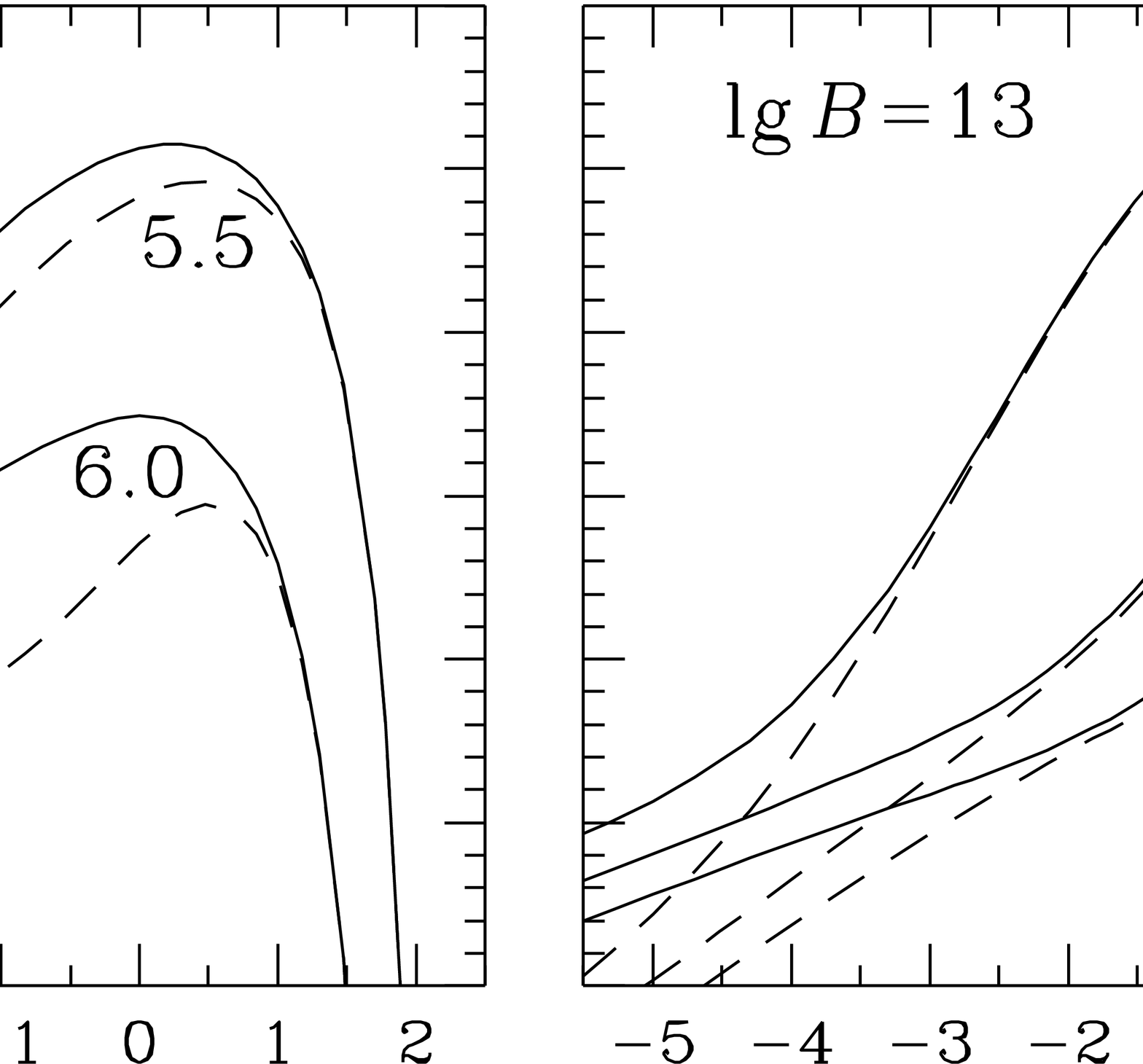}
\end{center}
\caption[ ]{\small{Fraction of atoms (solid lines --- total, dashed 
lines --- ground-state)
that contrubute to the opacities at different $B$ shown at the panels.
Numbers near curves are $\lg T$. At high $\rho$ neutrals  
dissociate due to the pressure ionization effect.}
\label{fig1}
}
\vspace{0.4cm}
\begin{center}
\leavevmode
\epsfxsize=5.cm
\epsfbox{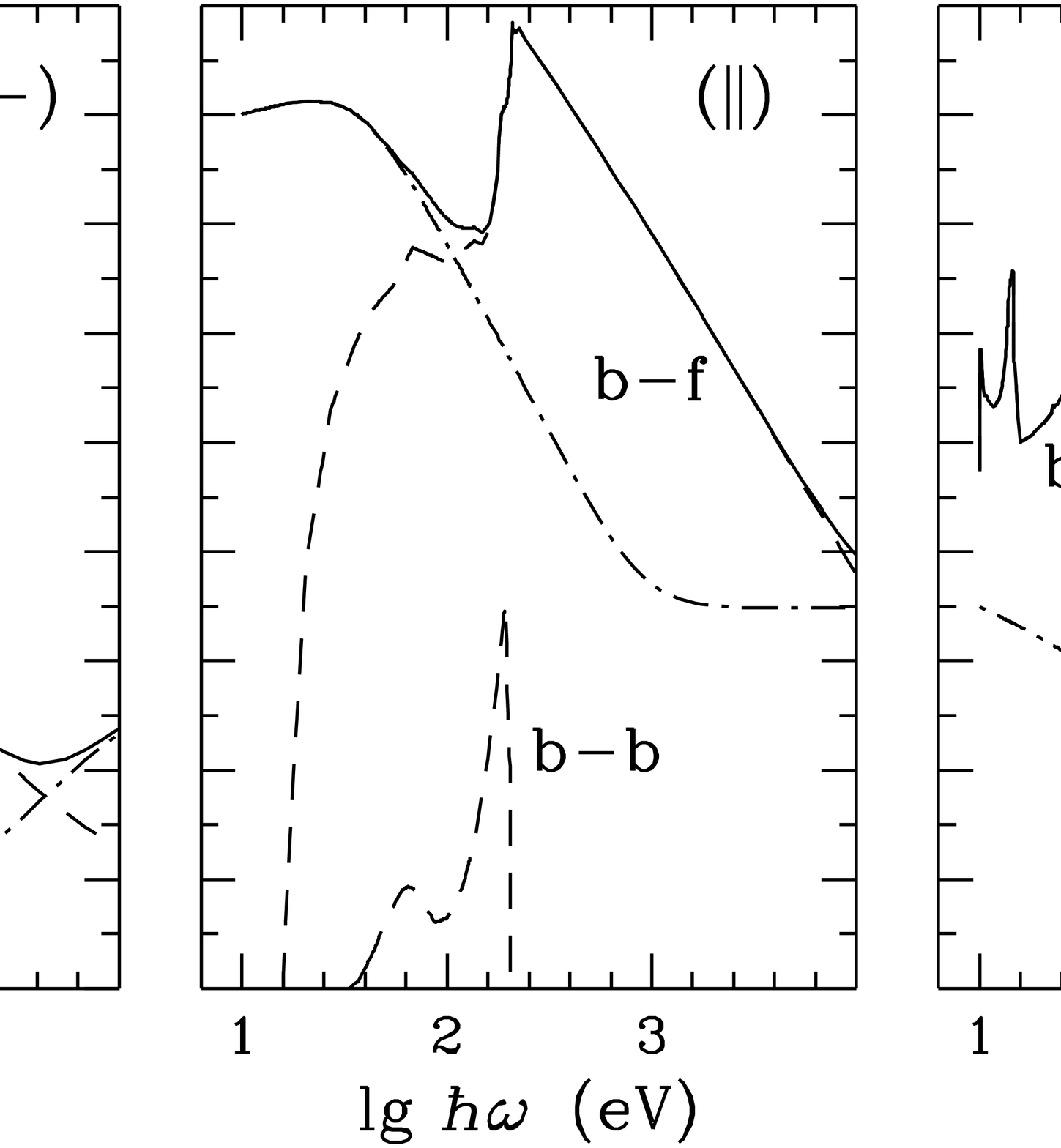}
\end{center}
\caption[ ]{\small{Opacities of hydrogen plasma at 
$B=2.35\times10^{12}$~G,
$T=3.16\times10^5$~K, and $\rho=0.1{\rm~g~cm}^{-3}$
for the radiation with
left (left panel), longitudinal (middle panel) and right (right panel)
polarization.
Dashed lines represent the contribution of ground-state atoms,
dot-dashed lines --- the opacity of the charged component,
and solid line --- the sum of both.}
\label{fig2}
}
\end{figure}

Figure~1 shows the neutral fraction $n_H^{\rm opt}/(n_H+n_p)$ that
contributes to the atomic opacities
   ($n_{s\nu}^{\rm opt}(K_\perp)\sim 
   n_{s\nu}(K_\perp)\exp[-4^3n_p v_{s\nu}(K_\perp)]$,
   cf. Potekhin, 1996a).
The total $n_H$ contributing to the EOS can be much larger in  
the low and high density limits. 
The different dependence   
of $n_H^{\rm opt}$ on $B$ at high and low $T$ is explained by 
the competition of  
change in $\chi_{s\nu}$ and the  
phase space occupied by electrons and atoms with $B$.   
The opacities shown in Fig.~2 are obtained using the bound-bound 
and bound-free cross sections calculated 
previously (Potekhin and Pavlov, 1997 and ref.~therein), 
and absorption coefficient of the ionized  
 fraction calculated following Shibanov et al. (1992).
The moving bound species contribute 
largely in the opacities of
typical cooling NS atmospheres, 
their spectra  are considerably 
different from those of the static-atom approximation 
(cf. Pavlov et al., 1995).    
This impels us to revise the previous 
atmosphere models at $T~\lsim~10^6$~K.
To do that properly one 
has to include the  
absorption by excited atoms, which can be substantial   
at relatively high $T\sim10^6$~K (see Fig.~1), 
and the bound-quasifree transitions (cf. Stehl\'e and Jacuemot, 1993) 
which are expected to fill the gap  
seen in the absorption for circular polarizations between b-b and 
b-f contributions.

This work was supported by RFBR grants 96--02--16870a and 96-07-89305,  
DFG--RFBR grant 96-02-00177G, and INTAS grants 94-3834 and 96-0542. YS 
thanks the LOC of the ``NSs \& Pulsars'' for the support during the meeting.
\vspace{1pc}
\re
Becker, W. and Tr\"umper, J., 1997, A\&A, 326, 682
\re
Blinnikov, S.~I., Dunina-Barkovskaya, N.~V. and Nadyozhin D.~K., 1996,
ApJS, 106, 171
\re
Caraveo, P., Bignami, G.~F. and Tr\"umper, J., 1996,
A\&A Rev., 7, 209
\re
Chabrier, G., 1990, J.~Phys. (Paris), 51, 1607
\re
Chabrier, G. and Potekhin, A.~Y., 1998, in preparation
\re
Heyl, J.~S. and Hernquist, L., 1997,
ApJ, 489, L67
\re
Lai, D. and Salpeter, E.~E., 1997, ApJ, 491, 270
\re
Page, D., Shibanov, Yu.~A. and Zavlin, V.~E., 1996,
in 
H.~U. Zimmerman, J.~Tr\"umper and H.~Yorke (eds.), 
MPE Report 263 (MPE, Garching) 173
\re 
Page, D., 1997,
ApJ, 479, L43
\re
Pavlov, G.~G. and M\'esz\'aros, P., 1993, ApJ, 416, 752
\re
Pavlov, G.~G., Shibanov, Yu.~A., Zavlin, V.~E. and Meyer, R.~D.,
 1995, in 
M.~A. Alpar, \"U.~Kizilo{\u g}lu and J.~van Paradijs (eds.), 
The Lives of the Neutron Stars, NATO {\it ASI\/} Ser.\,C, 
v.\,450 (Kluwer, Dordrecht) 71
\re
Pavlov, G.~G. and Zavlin, V.~E., this volume
\re
Potekhin, A.~Y., 1996a, Phys.\ Plasmas, 3, 4156
\re
Potekhin, A.~Y., 1996b, A\&A, 306, 999
\re
Potekhin, A.~Y. and Pavlov, G.~G., 1997, ApJ, 483, 414
\re
Potekhin, A.~Y., 1998,
J.\ Phys.\ B, 31, 49
\re
Rajagopal, M. and Romani, R., 1996,
ApJ, 461, 327
\re
Shibanov, Yu.~A., Zavlin, V.~E., Pavlov, G.~G., Ventura, J., 1992,
A\&A, 266, 313
 \re
 Stehl\'e, C. and Jacquemot, S., 1993, A\&A, 271, 348
\re
Vasisht, G., Kulkarni, S.~R., Anderson, S.~B., Hamilton, T.~T. and Kawai N.
1997, ApJ, 476, L43
\re
Ventura, J., Herold, H., Ruder, H. and Geyer, F., 1992,
A\&A, 261, 235
\re
Walter, F.~M., Wolk, S.~J. and Neuh\"auser, R., 1996, 
Nature, 379, 233
\re
Zavlin, V.~E. and Pavlov, G.~G., 1998, A\&A, 329, 583
\re
Zavlin, V.~E., Pavlov, G.~G. and Tr\"umper, J., 1998, A\&A, 331, 821
\end{document}